# End-coupled random lasers: a basis for artificial neural networks

**Niccolò Caselli[1][†][*], Antonio Consoli[1,2][#], Ángel María Mateos[1], Cefe López[1][§]**

**Light interference in strongly disordered photonic media can generate lasers where random modes are amplified in unpredictable way. The ease of fabrication, along with their low coherence caused by multiple small-linewidth peaks, made random lasers (RL) emerging, efficient, speckle-free light sources and a means to achieve spectral super-resolution. With potential to become a mature and accessible technology, their complex system's nature furnishes endless opportunities to unveil fundamental physics, since they can act as elements of optical network architectures. To date no experimental studies have analyzed the optical interaction between independent resonators in networks of RLs. Realizing RLs with a pumped strip joining two rough mirrors, we experimentally investigate and numerically simulate the emergence of networks when, by sharing scattering centers, RLs become coupled. We prove that the emission of a single RL can be manipulated by the action of others in the network, giving rise to substantial peak rearrangements and energy redistribution, fingerprint of mode coupling. Our findings, involving a few coupled RLs, firmly set the basis for the study of full-grown photonic networks. Oddly, both their deep understanding and their deployment as hardware clearly point in the direction of a novel disruptive technology: artificial intelligence on photonic random neural networks.**

## INTRODUCTION

Artificial intelligence, the kind of activity certain machines do to react to and thrive in their environment, is very often embodied by machine learning where computing machines run algorithms that improve through experience (1). Within this scheme deep learning is an approach, vaguely inspired in the animal brain, in which networks comprising multiple layers of interconnected agents called neurons, carry out a learning processes that enables them to solve situations never experienced previously. To date, the vast majority of studies in AI are still theoretical and algorithmical and applications, carried out in ordinary computers, have proved extremely powerful in tasks such as image and speech recognition for instance. But ordinary computers (relying on the von Neumann architecture) are not the best fit for the massively parallel processing required and call for architectures adapted to non-sequential processing.

Artificial neural networks can be realized as dedicated implementations where the computing process is performed by the collective interactions between the physical elements of the network. Among many optical implementations (2), nanophotonics has gained attention for mimicking both structural and functional features of neural networks by developing integrated optical circuits (3)and coupled lasers where the non-linearity and chaotic behavior are given by the amplified light emission (4–6). Non-linearity is an essential feature in the functioning of artificial neural networks. It is at the heart of many physical phenomena such as synchronization in coupled systems (7) and offers opportunities for application in complex networks(8) and chaotic systems (9), spanning oscillatory mechanical modes (10), spatiotemporal chaos (6), fluid dynamics (11), or synaptic neurons (12).

However, hardware based on lasers almost solely contemplates Fabry-Perot types of lasers, where the emission spectrum comprises a single emission line(13–15). Conversely RLs are intrinsically multiline and their architectures can easily give rise to complex networks(16–18). These features add to the natural non-linearity an additional spectral dimension which enriches the interaction and makes RL networks an ideal artificial neural networks platform.

At variance with conventional lasers, RLs do not require the engineering of a high-quality cavity, since they are based on disordered media in which the feedback necessary to trigger the laser amplification is provided by strong elastic light scattering(19–21). Therefore, they demand immensely easier manufacture, which allows inexpensive devices. A RL typically presents many narrow-line modes that cannot be prepared *a priori* as in a conventional cavity laser. The resulting emission shows lower spatial coherence, furnishing speckle-free illumination[22], as well as enabling spectral super-resolution[23]. RLs and, more generally, optical random modes, have found application in a large variety of fields from energy harvesting[24], information processing[25], bright terahertz emitters[26] to sensing[27] and tumor cell recognition[28], to name a few.

[1]Instituto de Ciencia de Materiales de Madrid (ICMM), Consejo Superior de Investigaciones Científicas (CSIC), Calle Sor Juana Inés de la Cruz, 3, 28049 Madrid, Spain. [2]ETSI de Telecomunicación, Universidad Rey Juan Carlos, Calle Tulipán, 28933 Madrid, Spain. [†]Current address: Departamento de Química Física, Universidad Complutense de Madrid, Avenida Complutense, 28040 Madrid, Spain. [*]n.caselli@csic.es, [#]antonio.consoli@urjc.es, [§]c.lopez@csic.es

The most widespread RL implementation is based on distributed feedback, where the scattering centers are randomly placed inside the optically active medium. We previously introduced an alternative configuration in which amplification and feedback are separated and performed by a narrow strip pumped region and two rough mirrors respectively. The later provide feedback in a way similar to a Fabry-Perot cavity but by virtue of the disorder embedded in the mirrors, the emission exhibits RL behavior (29, 30). Since their geometry remains static, they show stationary emission spectra allowing to disentangle the interactions when single RLs are engaged in a network.

Coupling between modes has been studied in a single RL (30, 31) but not between multiple RLs. Here, we demonstrate that this modal interaction can be induced in multiple disordered resonators, giving rise to a new class of active optical networks based on interconnected and effectively coupled RLs. The network, defined in the plane of the gain material, is determined by several, distant, nondescript, scattering centers (rough mirrors) multiply connected by pumped strips. The system can be considered as a unique collective resonator in which manifold coupled oscillators are turned on at will (by turning on the pumping strip). This is proved by the fact that highly correlated spectral signatures are observed at all the nodes of the network.

Numerical calculations based on coupled mode theory were carried out and predictions on the spectral redistribution induced by the interacting RL network were found in agreement with the observed synchronization behaviors. Finally, since a single network node can be used as scattering and coupling element between many resonators, more complex network architectures can be envisioned as "join-the-dots" graphs in order to extend the proposed approach to simulate neural networks even with the ability to use any node as output.

## NETWORK PREPARATION

For the sake of simplicity, before extending the study to more complex geometries, we demonstrate the functionality by focusing on the basic building block of a network: two equivalent RLs connected through a node, consisting of a scattering center that furnishes feedback for both random resonators to lase. In this configuration, we prove that one RL emission can be affected by the action of the second RL with which it has a scattering center in common. Then we realized different network archetypes involving more than two mutually linked RLs with increasing complexity and proved that every investigated arrangement can sustain RL interactions.

At variance with ordinary cavity lasers, RLs exhibit a multitude of modes engaged in a complex interaction of gain and loss, virtually impossible to foresee for the intrinsic difficulty in obtaining useful prior knowledge of the actual cavity and by the complex nature of the mode competition.

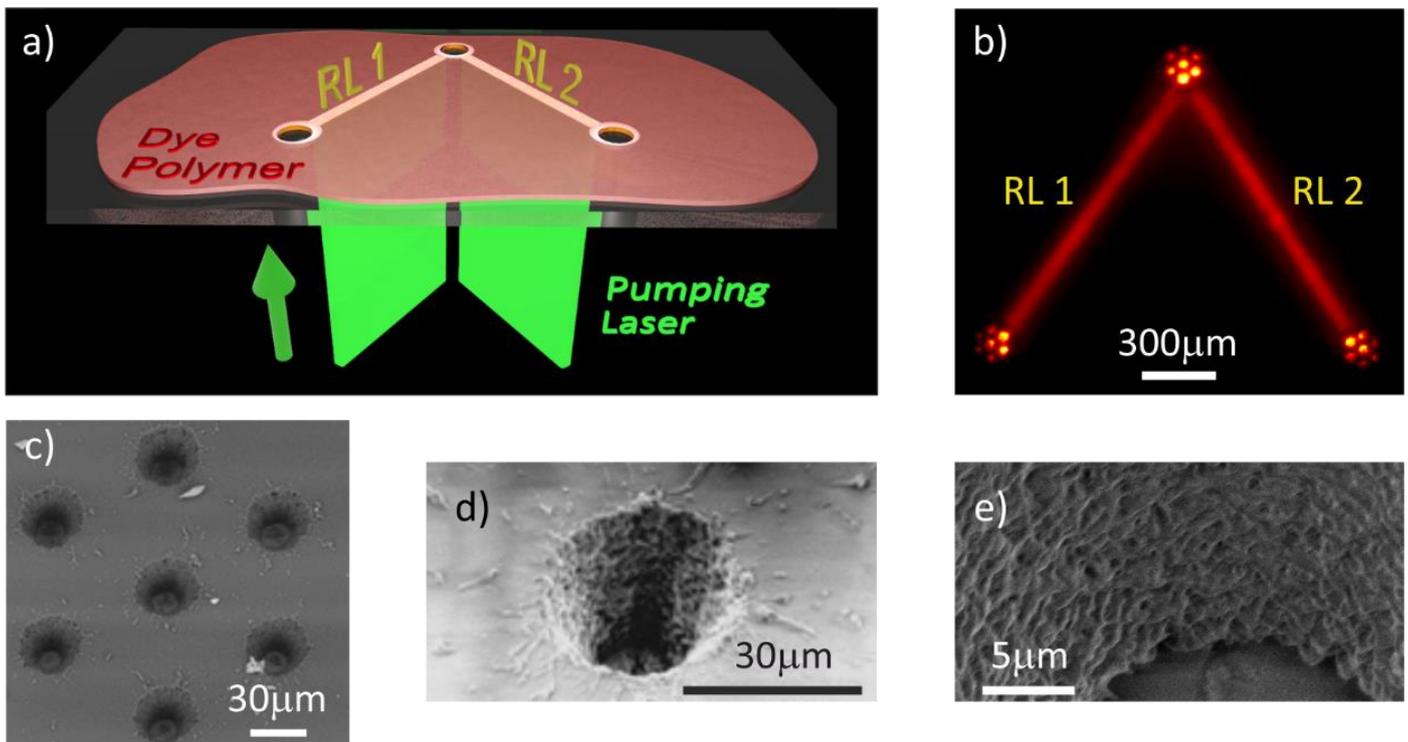

**Figure 1.| RL network fabrication** a) Schematics of the RLs network. The dye-doped polymer thin film is deposited on a glass substrate and scattering centres are created by drilling the film. Optical excitation is performed by a spatial light modulator sculpted laser impinging from the bottom and uniformly pumping regions that connect the scatterers. b) Emission intensity image (acquired by CCD) of two RLs pumped simultaneously. Each resonator consists of a rectangular pumped volume placed between two scattering centres each made of seven holes with disordered internal surfaces. c-e) Scanning electron microscopy (SEM) images of one of the scattering centres in b), reported with increasing magnification to highlight, inside a given hole, the rough sidewalls, which are responsible for the RL feedback action.

The typical single RL we used, representing the building-block of RL networks, is composed of two distant ($\approx 2$ mm) scattering centers acting as rough mirrors and connected by a laser pumped stripe[30]. When optical gain overcomes losses, amplified spontaneous emission, which is spectrally coupled with modes of the unconventional cavity, generates a RL. The active medium in which the random lasing occurs is a solid, dye-doped biopolymer thin film (see Methods) in which the polymer matrix has been chosen for its ability to prevent dye quenching[32, 33]. The networks of RLs are realized by inscribing scattering centers by means of a direct laser-writing technique into the polymer film and by optically pumping the dye-polymer region through an ensemble of stripes projected on the sample surface (see the schematics of Fig. 1a, the out-of-plane emission shown in Fig. 1b and Methods).

Each scattering center consists of a single (Fig. 1d,e) or a series of holes (diameter size, tens of μm), as those -arranged in a hexagonal pattern- reported in Fig. 1b-c. The geometrical arrangement obeys only practical reasons and no advantage is derived from their regularity beyond the economy of manufacture. The rough internal surfaces of the drilled holes form a disordered air/polymer interface whose mission is to strongly scatter light coming from the pumped volume (see Fig. 1b), thus providing the random feedback for lasing action. The number of laser modes sustained by a RL, as well as their threshold and angular distribution, has been proven to depend on the roughness and porosity of the scattering surfaces[34, 35].

The scattering centers act as disordered mirrors and, at the same time, as out-of-plane couplers, which allow to probe the lasing modes by recording the scattered emission orthogonally to the sample surface (see Methods). It is worth mentioning that random modes in this kind of RL emerge as stable sharp random spectral peaks and the same resonances are recorded in both ends (see Fig. S1, S2). This indicates that the device consisting of a pumping line and two scattering centers is acting as a single oscillator, supporting modes some of which enjoy sufficient gain to be amplified in the round-trip between the disordered mirrors[29].

## MEASURING THE COUPLING

When considering RL networks, one must establish a protocol to determine the existence of coupling on the basis of the spectra acquired from each or the compound RL. Two different RLs provide spectra that can be taken as standard for independence (fully uncorrelated) whereas two spectra taken successively from a single RL (only subject to stochastic fluctuations) can be used as standard for strong dependence and similarity, since high correlation is assured.

We started by addressing the simplest case consisting of two RLs having one scattering node in common. In this configuration, two resonators activated one at a time show distinct, uncorrelated spectra that we take as benchmark of spectral independence. When both RLs are pumped simultaneously, if they do interact (coupling through the common disordered mirror) the emerging spectral features must depend on the lasing state of both resonators. Several circumstances can be expected to occur such as frequencies common to both uncoupled resonators not belonging to the compound one, modes from some resonator becoming available to the compound one, modes not lasing in any resonator showing as a consequence of the coupling and, generally, changes in mode gain. This provides an alternative method to detect coupling between resonators: searching for dissimilarity between the spectrum of one when pumped alone and when in conjunction with the other. The fingerprint of an effective interaction would thus be a spectral modification, *i.e.* peak wavelength-shifts and/or an intensity redistribution, that may be detected at every network node. Therefore, the *modus operandi* we followed to test mode coupling occurrence was acquiring spectra of each single RL and compared

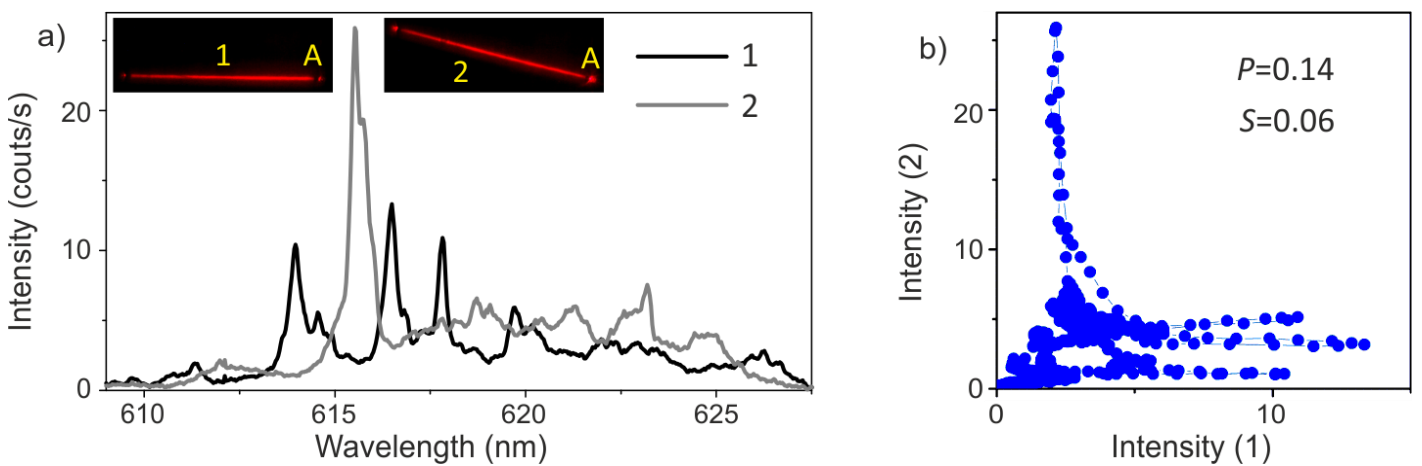

**Figure 2. | Independent spectra from two RL** with scattering centres made of single holes of 30 μm diameter size as an example of fully uncorrelated spectra. a) Spectrum acquired in position A by pumping line 1 (black curve) and line 2 (grey curve), respectively. In the insets the out-of-plane emissions of the two cases are reported. Both lines length is 2 mm. b) Parametric plot of the intensity of the spectra reported in a). The values of Pearson correlation (*P*) and of the area of the parametric plot (*S*) are reported for each case.

them to the spectrum emerging from the simultaneous activation of both RLs in the network.

To set a standard for uncoupled RLs we first studied the device reported in the inset of Fig. 2a, where two RLs (labelled 1 and 2) share the scattering center A. The device is composed by three scattering nodes made of single holes connected by two pumping lines. The two RL emissions (spectra 1 and 2) are expected to be independent, as they originate from two different resonators non-simultaneously activated and, as shown in Fig. 2a, they show two uncorrelated sets of peaks.

In order to visually highlight the low correlation and spectral difference between the two RLs, in Fig. 2b we reported the intensity of the first spectrum as a function of the intensity of the second one, for every acquired wavelength: each data point having coordinates $[I_1(\lambda_i), I_2(\lambda_i)]$ for $i = 1 \ldots n$, with $n$ corresponding to the number of wavelength data. This representation will be hereafter called parametric plot. Since in this picture the vast majority of data lay outside the diagonal line, a low linear correlation is evident, as quantified by the small value of the Pearson coefficient being $P = 0.14$[36, 37]. The high intensity peaks of one RL lying on background intensity regions in the other RL, appear as vertical and horizontal streaks, while data in fluorescence background regions of both resonators appear as a cloud at lower intensities. Obviously two proportional spectra would give a straight line in this kind of representation, with the Pearson correlation directly proportional to the regression coefficient.

An additional measure to assess the degree of similarity between spectra, can be provided by the surface area (S) enclosed by the lines that connect the parametric plot data. The surface area measure increases with shifted, overlapping peaks but, at variance with Pearson's correlation it attains very small values both for high correlation (e.g. between nearly proportional spectra, see Fig. S3, S4) and very low correlation (vertical and horizontal streaks in Fig. 2b). Therefore, a high $S$ value implies that the overlap is not complete, but that the spectra exhibit small variations in the peak positions and/or amplitudes, as occurs if an effective weak interaction was in place. In our analysis, it is the conjunction of the parameters $P$ and $S$ that allows to estimate the coupling between single RLs in the network. For instance, for the two independent RLs reported in Fig. 2a the low value of both parameters ($P = 0.14$ and $S = 0.06$) proves the negligible correlation between them.

When two resonators do couple, this interaction must appear as a difference between the sum of their independent spectra and the compound spectrum obtained with simultaneous pumping. We used the dissimilarity between these spectra as a smoking gun of coupling. When both RLs are simultaneously active, the compound spectrum (1&2), reported in Fig. 3a (red line), was acquired at the common scattering node, in order to overcome fluctuations in the out-of-plane emission between different nodes. This emission was compared to the sum of the separate spectra taken successively (1+2) at the same node, reported in Fig. 3a (blue line). If no interaction were taking place, each RL would independently scatter their emission out of the network and the compound emission would be the sum of the two single RL spectra, resulting in strong overlap and high correlation ($P \approx 1$; $S \approx 0$). The extreme opposite would lead to compound spectrum retaining no features from either resonator, thus bearing no resemblance with their sum, in which case a streaked parametric plot would result ($P \approx 0 \approx S$). A case where $P \approx 0$; $S \approx 1$ cannot occur for spiked spectra (or even background fluorescence) as it is a signature of fully random noise signal.

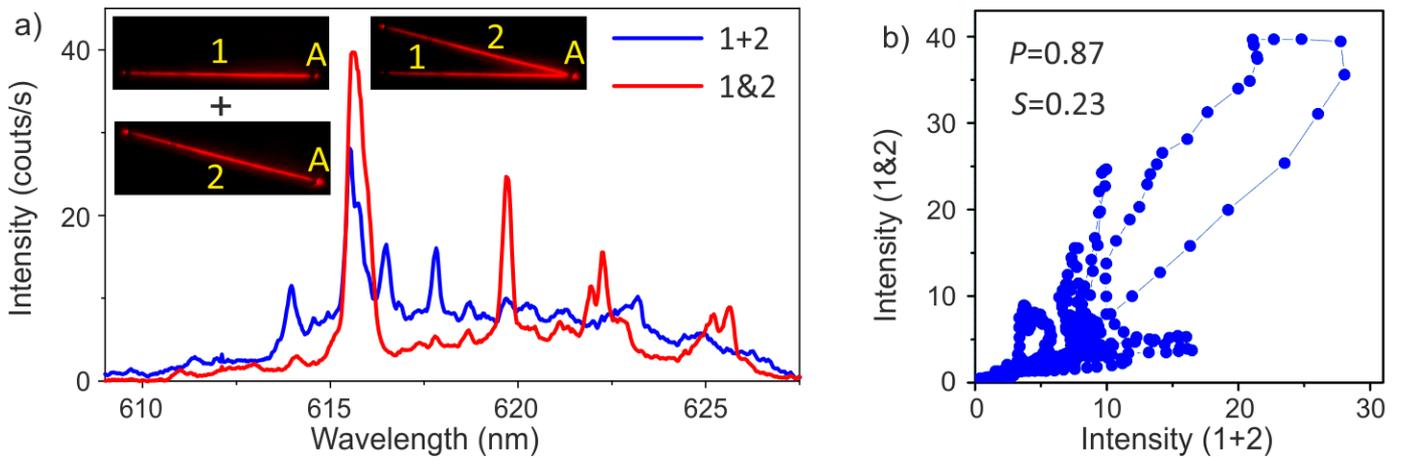

**Figure 3. | Two resonators RL network** with scattering centres made of single holes of 30 μm diameter size, corresponding to the one reported in Fig. 2. a) Spectra obtained by summing the single emissions (1+2, blue curve), and by pumping simultaneously both lines (1&2, red curve). All the spectra are detected in position A. b) Parametric plot of the intensity of the spectra reported in a). The values of Pearson correlation ($P$) and of the area of the parametric plot ($S$) are reported for each case.

In Fig. 3a-b the sum spectrum shows a clear difference with the emission detected from the compound resonator, exhibiting lessened correlation, $P \approx 0.87$, and a moderate area $S \approx 0.23$. Typical values for correlation in successive acquisitions of one resonator emission (See Fig. S2) provide a benchmark for highly correlated spectra as high as $P \approx 0.98$. A redistribution of the spectral density with suppression and/or enhancement of existing modes, up to the occurrence of new ones, is observed. It is remarkable how the compound emission is dominated by two strongly enhanced peaks at 615.6 nm (giving rise to the large loop in the parametric plot, responsible for most of the value of $S$) and 619.7 nm (originating one large vertical streak), while the majority of the emission is inhibited with respect to the sum spectrum. The emission stability of the RLs was checked in every spectral acquisition (see Method and Fig. S2) and the differences observed between the sum and the compound spectra cannot be ascribed to variations in the exciting conditions, polymer degradation or detection instabilities.

Investigation of 20 network devices with the same design revealed that the spectral rearrangement occurs with large variations, as depicted in Fig. S9. We found fingerprint of coupling ($S > 0.15$ and $P < 0.9$) in 40% of the cases. This inhomogeneity is the result of random variations in mode gain and coupling strength, which depends on the specific features of the resonators and on spectral and spatial overlap of RL modes, that cannot be engineered *a priori*. The analysis presented in this work aims to prove that individual RLs composing a network can exhibit mode coupling between them.

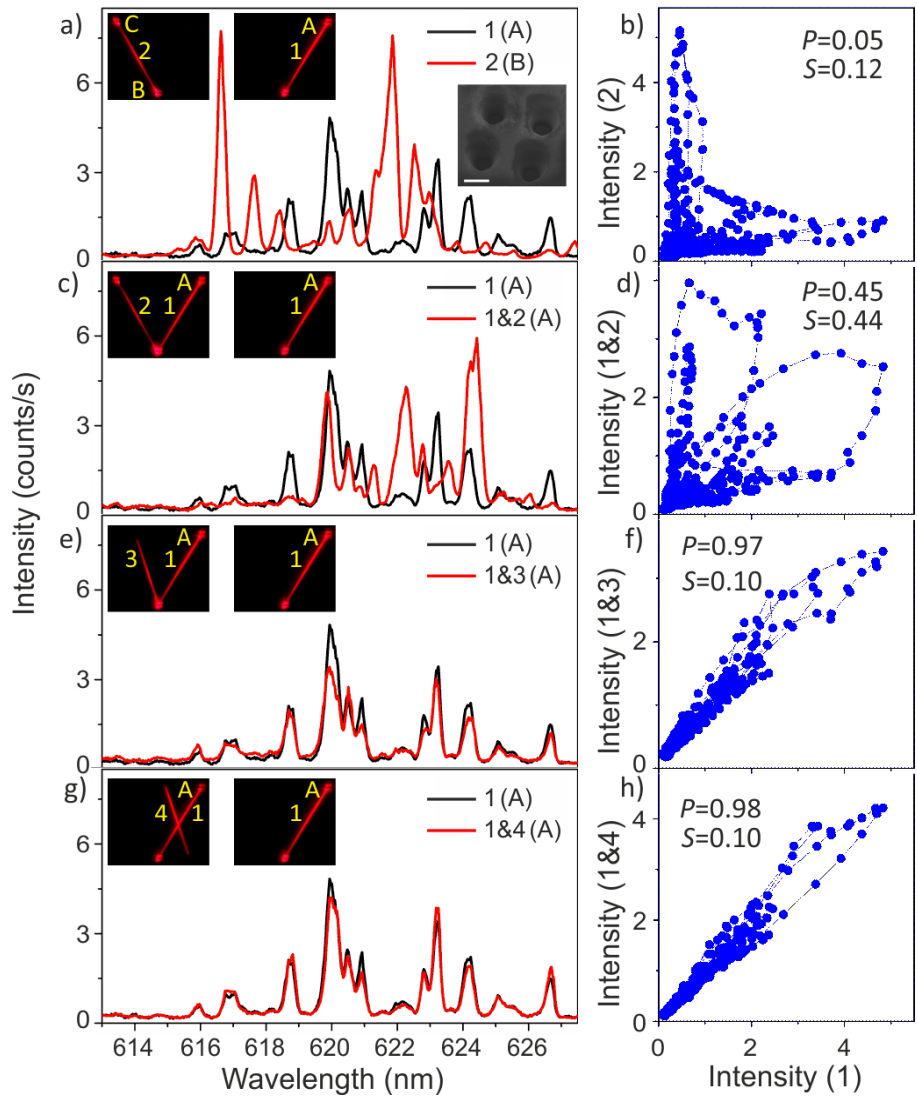

**Figure 4. | RL network** with scattering centres made of holes (50 μm diameter size) arranged in square pattern with 100 μm side. a) Spectrum acquired in position A by pumping line 1 (black curve) and in position B by pumping line 2 (red curve). In the insets the out-of-plane emission of the two RLs is reported, along with the SEM image of one scattering centre (scale bar 50 μm). b) Parametric plot of the intensity of the spectra reported in a). c) Spectra acquired in position A by pumping line 1 (black curve), and line 1 and 2 simultaneously (1&2, red curve). d) Parametric plot of the intensity of the spectra reported in c). e) Spectra obtained in A by pumping line 1 (black curve) and line 1 and 3 simultaneously (1&3, red curve). f) Parametric plot of the intensity of the spectra reported in e). g) Spectra obtained in A by pumping line 1 (black curve) and line 1 and 4 simultaneously (1&4, red curve). h) Parametric plot of the intensity of the spectra reported in g). The values of Pearson correlation (*P*) and of the area of the parametric plot (*S*) are reported for each case.

Apart from investigating the sum *vs* compound spectral similarity at the common node, another experimental approach worth studying to prove mode coupling is evaluating how one RL emission from nodes that are not shared is influenced by the network interaction. Moreover, to prove the universality of the coupling occurrence, we varied the geometry of the scattering centers by fabricating networks with three nodes, positioned at the corners of an equilateral triangle, each consisting of four holes arranged in a square pattern, as reported in the inset of Fig. 4a. Two single RLs are induced by pumping the lines that connect corners A-B (line 1) and B-C (line 2), respectively, as shown in Fig. 4a. The corresponding parametric plot demonstrates the low correlation between spectra from two independent single resonators excited successively ($P = 0.05$; $S = 0.12$). In order to test the influence of RL 2 on the emission of RL 1, spectra were collected from position A when RL 2 was (Fig. 4c red line) or was not pumped (Fig. 4c black line). In this position the emission of RL 2 alone is null in the absence of RL 1 action (see Fig. S6), so that, trivially, the sum 1+2 equals the emission of RL 1. When RL 2 was turned on, new modes got amplified through line 1 and were observed in node A, for instance those at 608 nm and 623 nm, which belong to the emission of RL 2. At the same time, some modes experienced a suppression, like those at 614 nm and 616 nm. The parametric plot, reported in Fig. 4d, furnishes a heavily reduced linear correlation,

$P = 0.45$, and a high area value, $S = 0.44$, thus bearing evidence of a strong interaction between the two RLs.

In order to close down the definition of end-coupled RL network we study the impact of a strip with only one disordered mirror (not a resonator) as line 3 in Fig. 4e, coupled to a full-fledged RL or one that only intersects the active region with RL 1 (line 4 in Fig. 4g). These cases are reported in Fig. 4e-h where the high correlation ($P \approx 0.97$) and low area parameter ($S \approx 0.1$) prove the lack of coupling. The latter case presents a pump line without ending mirrors but even if had both ending scattering centers the only intersection being a point in the gain strip precludes the inclusion in the network (see Fig. S7). This highlights the requirement to share a mirror for a laser to belong to the network. Further examples are reported in the Fig. S6.

## RING NETWORK

A ring RL network, still simple yet more complex than the one investigated in Fig. 4, was considered too. It consists of three disordered scattering centers each made of 7 holes with rough internal surface (see Fig. 1b-e) that can be joined by three pumping strips. Here, we exploited all three single RLs emerging by pumping the device with lines connecting the scattering centers. When all single RLs are pumped at the same time, the spectral fingerprint collected at each node position show high correlation (see Fig. S12), proving that the system is a single, compound cavity with modes that cannot be exclusively associated to a given RL. We focused on the effect induced by switching on/off a single RL of the network. In particular, the network emission variation induced by turning on the pumping line B-C (RL 3) is reported in Fig. 5a-b. The emission extracted from A (that does not belong to RL 3), was strongly modified by the action of the random modes of RL 3. The analysis of the parametric plot of the spectra collected when *i*) only lines 1 and 2 are pumped and *ii*) the three lines are pumped at the same time, gives $P = 0.74$ and $S = 0.32$, indicating a strong spectral modification that, again, produces a reduced correlation and a high enclosing area. In this case the coupling fingerprints are the spikiness of the spectra causing small wavelength shifts and large intensity variations, which contribute to large loops in the parametric plot. A similar behavior, although resulting in a weaker interaction ($P = 0.71$ and $S = 0.19$), is found by measuring in C the effect of switching RL 2, reported in Fig. 5c-d. Comparison between the three scattering centers morphologies presented in this work shows that larger hole number, i.e. increased disorder surfaces, boost the chances to establish coupling.

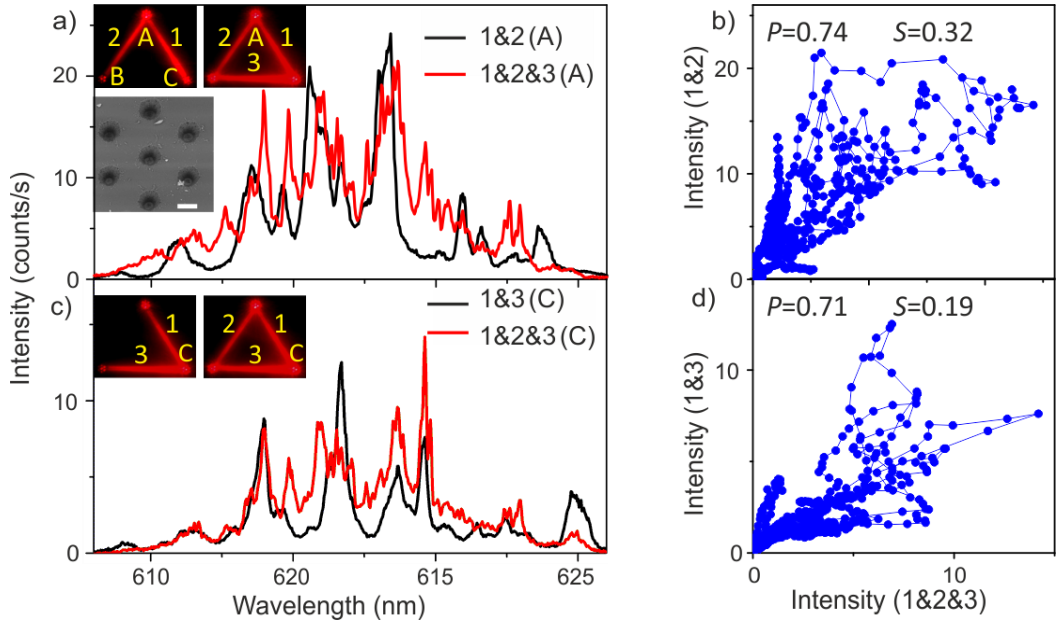

**Figure 5. | Ring RL network** with scattering centers made of holes (diameter size of 30 μm) arranged in hexagonal pattern. a) Spectrum acquired in position A by pumping lines 1 and 2 (1&2, black curve), and by pumping all the lines at the same time (1&2&3, red curve). In the insets the corresponding out-of-plane emissions are reported, along with the SEM image of one scattering center (scale bar 30 μm). b) Parametric plot of the intensity of the spectra reported in a). c) Spectra acquired in position C by pumping lines 1 and 3 (1&3, black curve), and by pumping all the lines (1&2&3, red curve). d) Parametric plot of the intensity of the spectra reported in c). The values of Pearson correlation (*P*) and of the area of the parametric plot (*S*) are reported for each case.

The analysis based on the conjunction of the Pearson's correlation and the enclosed area allowed to quantify the coupling in a network of RLs, since the interaction induces a spectral redistribution of the amplified modes. With respect to other method that evaluate the similarity between different spectra, such as the autocorrelation function([38]) or the mutual information([39]) our research gives a remarkable outcome in highlighting spectral redistributions (see Fig. S9).

### NUMERICAL MODEL
In order to substantiate the experimental findings we simulated the RLs network in the framework of time coupled mode theory([40, 41]). For each mode only the temporal and spectral evolution were considered (no spatial information was included). We introduced mode gain, losses, coupling and also gain saturation, while, for sake of simplicity, we neglected other second order non-linear effects. We created a large pool of 40 modes, from which a RL was assigned 10 modes ($j,k = 1...10$, randomly chosen among the globally defined, see Methods), which satisfy a time dependent set of differential equations:

$$\dot{a_k} = i\omega_k a_k - \alpha_k a_k + \sum_{j \neq k} c_{j,k} a_j + g(t, \omega_k)\frac{a_k}{1 + \gamma_k |a_k|^2} \quad (1)$$

where $a_k(t)$ is the complex amplitude of the $k$-th mode as a function of time, $\omega_k$ the frequency, $\alpha_k$ the losses, $c_{j,k} \propto |\omega_k - \omega_j|$ the coupling coefficient with all other $j$-th modes that fade with detuning, $g(t, \omega_k)$ the time dependent mode gain (pump pulse) and $\gamma_k$ the gain saturation coefficient. Eq. (1) describes how the $k$-th mode evolves in time. By Fourier transforming the solution, we obtained the spectral profile in amplitude and phase of each mode and the full RL spectrum is retrieved by summing the intensity of all modes. In this way, two independent RLs are represented by two sets of 10 modes randomly chosen from the pool, for which Eq. (1) is solved separately, that is, couplings are only considered within the 10-sets giving rise to the grey and black spectra reported in Fig. 6a, respectively. Notice that, due to the coupling, not all modes necessarily merged in the spectrum, having merged with neighboring ones. The parametric plot of Fig. 6b, heavily dominated by vertical and horizontal streaks, highlights the very low correlation between them.

In modelling the case of two coupled RLs, Eq. (1) was solved considering that the two 10-sets of modes owned by each RL

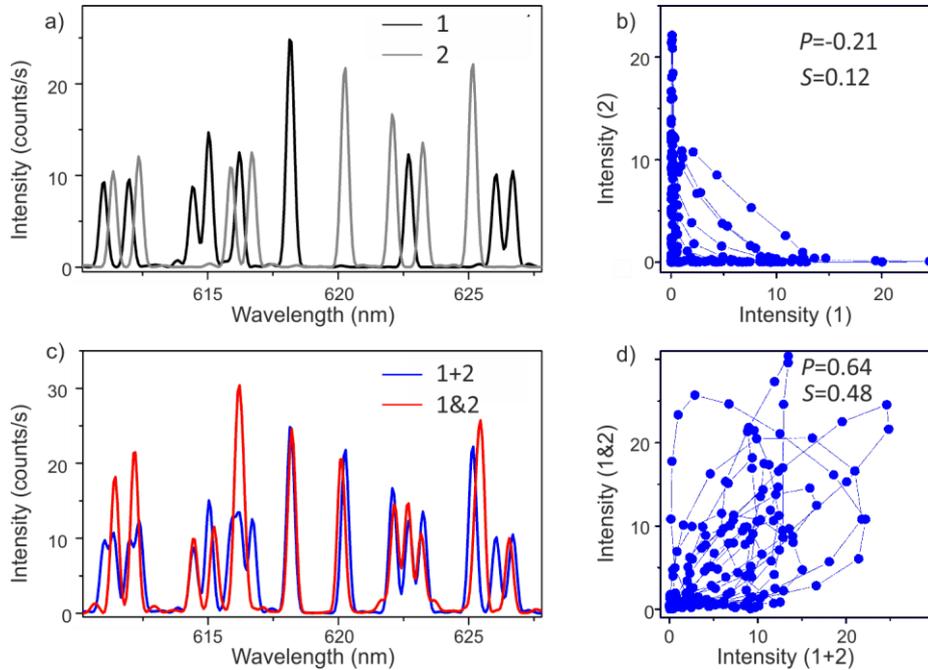

**Figure 6. |Numerical simulation** employing coupled mode theory. a)-b) Spectra corresponding to two independent RLs, black and grey curves in a), show a low correlation in the parametric plot in b). c) The sum of the spectra in a) (1+2, blue line) is compared to the spectrum (1&2, red line) obtained by introducing a coupling term between the two sets of modes. d) Parametric plot of the spectra in c). The value of Pearson correlation (*P*) and of area parameter (*S*) are reported for each case. The coupling gave rise to an intermediate value of *P* and a high area parameter, which therefore can be used as experimental evidence of RLs interaction.

contributed in the equation, as they were pumped at the same time. In this case, modes belonging to different sets were randomly interleaved and subject to coupling. The resulting compound spectrum, 1&2, (red line in Fig 6c) was compared to the sum of the single resonators, 1+2 (blue line in Fig. 6c). The coupling gave rise to a sizable spectral redistribution with respect to the sum of the stand-alone emissions. In fact, the frequency and intensity of the peaks are slightly changed, with attraction/repulsion between adjacent modes and energy transfer due to the mode interaction. The corresponding parametric plot of Fig. 6d, resembles the experimental observations reported in Fig. 3d, Fig. 4b and Fig. 5d by showing a linear correlation $P = 0.64$ along with a high area parameter $S = 0.48$. This demonstrates that in the presence of an effective interaction, the spectra show a reduced linear correlation along with a high area value, as we found experimentally. By averaging the results of 20 simulations that compare (1+2) to (1&2) with different sets of independent modes and coupling strength, we found consistent behaviors (see Fig. S10).

## DISCUSSION
We realized multiple RLs in a planar morphology by inscribing scattering centers in a film of gain material made of dye-doped biopolymer and selectively pumping the strip joining them. RL resonators can be realized by using ablated holes in the active medium that act as rough mirrors by back scattering radiation into the pumped stripe and facilitating amplification. At the same time these scattering centers forward scatter so that they establish the coupling with any resonators sharing the center. Additionally, out of plane scattering helps monitor their lasing action by spectroscopic means.

By designing networks in which RLs (each consisting of a strip between two rough mirrors) share mirrors, we proved that it is possible to induce an effective coupling between them. Spatially resolved probing and spectral correlation analysis allowed to detect their synchronous ensemble interaction.

Tests of different configurations consistently show the network coupling effects when RL resonators are fully formed (pumped strip and two end rough mirrors) and the linking includes them in the network (at least one of the mirrors is shared).

Our findings are supported by coupled mode theory calculations that predicted the observed spectral correlations. The modelling, based on time coupled mode theory, gives a full qualitative account of the behavior despite dispensing with the full details of the actual cavities.

The presented architecture has a small footprint, low fabrication complexity and cost, making it a good candidate to study highly nonlinear interactions in networks. Spatially modulated pumping, where different RLs in the network can be selectively pumped (digital) or different pulse power sent to each of them (analog), enables a versatile random neural network platform. We believe that the ample potential for large connectivity and dynamic character of the pulsed action offered by this platform will trigger the study of more complex networks based on RLs aiming to realize photonic neural network architectures. Spec-tral synthesis achieved by training the network to produce desired peaks; reservoir computing configuration, to save testing large numbers of weights in training or spiked implementation, where the pulsed character of the RLs is exploited, are potential direct applications of this kind of platforms. Finally, this can be an initial step in developing statistical methods based on machine learning in order to address and evaluate the RL interaction even in more complex network geometries.

## METHODS
### Sample preparation
The dye-doped biopolymer films were produced by introducing 4-(dicyanomethylene)-2-methyl-6-(4-dimethylaminostyryl)-4H-pyran (DCM) molecules, with a 0.35% mass concentration in a deoxyribonucleic acid (DNA)-cetyltrimethyl ammonium (CTMA) polymer matrix. The cationic surfactant CTMA (Sigma-Aldrich 292737) was used to make the DNA (Sigma-Aldrich D1626) soluble in organic solvents and then easily mixable with the DCM dye-molecules solution. The DNA-CTMA complex is dissolved (4 vol.%) in ethanol and mixed with the DCM solution (0.5 vol.% in equal parts of ethanol and chloroform). The resulting blend was magnetically stirred for 5 hours. Then it was spread by dropcasting onto a glass substrate, and solid films with a thickness of about 20 μm were obtained after drying the sample at room temperature and pressure.

### Rough mirrors.
The scattering centers that serve as disordered mirrors and provide feedback for RL were fabricated by using a direct laser-writing technique. The polymer was locally ablated by means of a short-pulsed (100 fs), high-energy (~250 μJ/pulse) Ti:Sapphire laser emitting at 800 nm peak wavelength. The laser power, the number of shots delivered and the focusing position were controlled by using a half-wave plate in conjunction with a linear polarizer, a software controlling the laser output and motorized translational stages, respectively. A single laser shot focused by a lens of 5 cm focal length drilled a hole in the polymer matrix. In order to create a reproducible hole size, we overcame the shot-to-shot laser power fluctuations by lowering power and increasing the number of pulses (50 single shots) delivered to obtain each hole. This assures that, in the energy range between 200-300 μJ/pulse, the local ablation removed completely the polymer and reached the underlying glass substrate, thus leaving a cylindrical-like air-defect in the polymer matrix. By varying the laser fluence we achieved reproducible holes with diameter size in the range 30-150 μm.

### Random lasers measurements
The excitation setup involves a 10 ns pulsed, frequency doubled 532 nm Nd:YAG laser with a 10 Hz repetition rate, whose beam was shaped by an amplitude spatial light modulator working in reflection. By means of this scheme we imaged on the sample surface the intensity distribution required to produce the desired optical pumping geometry. We drew a set of stripes of same length and width (50-100 μm) connecting scattering centers, as reported in Fig. 1a-b. The out-of-plane lasing emission was imaged with a magnification equal to 2 both on a CCD, see Fig. 1b, and on a plane where an optical fiber tip connected to a spectrometer collected the signal, which was coupled to a spectrometer and a visible light detector (SPEC, Andor Shamrock 303). We achieved a spatial and spectral resolution of 50 µm and 0.1 nm, respectively. The fiber tip was mounted on motorized translation stages in order to collect the radiation emerging from different scattering centers. It is important to stress that by applying this pumping scheme, and by triggering the spectral acquisition to the laser repetition rate, consecutive spectra are independent from one another, since they refer to different excitation pulses. In order to prove the stability during the entire acquisition time, we repeated a sequence of illumination frames, each of which contains a different pumping lines configuration in the network. For instance, referring to Fig. 2 the sequence was composed by three frames: line 1, line 2, lines 1&2. For each frame we acquired 50 consecutive spectra, corresponding to 50 laser single shots, to be averaged. The sequence was repeated many times. Therefore, for each illumination geometry we achieved a series of nominally identical spectra mixed in time. In every series the spectral fluctuations were minimal, exhibiting a high degree of correlation ($P = 0.98$ on average, see SI) over a period of about 5 min. After longer times of constant exposure above laser threshold, the polymer matrix suffered modifications due to the induced heating, hence causing the spectrum to slowly change and to experience a correlation decrease with respect to the initial one that usually led to a complete decorrelation after 15 min (see SI). On grounds of such stability analysis, for every investigated RL configuration we performed a series of ten nominally identical acquisitions, averaged to obtain a reliable spectral signature. More details concerning pumping threshold, integration time and stability can be found in the SI.

**Numerical model.**

The system under investigation was numerically replicated by a set of $N = 40$ random possible lasing frequencies, almost evenly spaced (a 5% degree of fluctuation from equispaced was added) in order to cover the spectral range of interest. The excitation pulse was designed to reproduce the pump used in experiments with about 10 ns and a gain bandwidth of about 30 nm. The gain saturation coefficient $\gamma_k$ was considered equal for all modes. The coupling coefficient matrix $c_{j,k}$ in Eq. (1) was defined for all modes by assigning decreasing real values for increasing mode spectral detuning, with a linear slope. The maximum value of the coupling was chosen as $c_{max} = 1.5 \cdot 10^{-5}$ ($g_{max} - \alpha_{max}$). The same coupling matrix was used for calculating the spectra of the single resonators and of the coupled compound where all the modes interact simultaneously through $c_{j,k}$. In order to simulate the experimental lasing spectra, we considered that in each single RL 10 different modes (randomly selected among the $N$ possible) are allowed to lase. The introduction of mode coupling implies that the instantaneous frequency of the $k$-th mode, defined as the time derivative of the phase, varies with respect to its initial value, $\omega_k$ as a function of time, depending on the coupling to other modes. This effect explains the frequency shift from $\omega_k$, so that adjacent modes can partially overlap until, in the limit of high coupling, a single peak emission is found. In the framework of coupled mode theory, the peak modification is due to energy exchange in time between interacting modes.

## SUPPLEMENTARY MATERIALS

The data that support the plots within this paper and other findings of this study are available from the corresponding authors upon reasonable request.
The MATLAB codes developed to execute the calculations presented in this paper are available from the corresponding authors upon reasonable request.

**Acknowledgements**. This work was partially funded by the Spanish MCIU RTI2018-093921-B-C41 project. N. C. acknowledges MCIU *Juan de la Cierva* program. Álvaro Blanco and Francesco Riboli are thankfully recognized for fruitful discussions.

# "End-coupled random lasers: a basis for artificial neural networks"

Niccolò Caselli[1†]*, Antonio Consoli[1,2#], Angel Maria Mateos[1], Cefe López[1§]

## 1. Single device emission and stability

For a single RL consisting of two disordered scattering centers connected by a pumping line in the dye-active polymer, we acquired spectra in position corresponding to both scattering centers, labelled as A and B in the inset of Fig. S1a). They show sharp RL peaks only if the pumping energy of the pulsed laser overcomes a certain threshold. In our experiment the pump energy can be changed by acting on the spatial light modulator, that is the optical element that allows to draw any desired pumping laser geometry and intensity distribution on the plane where the sample is fixed.

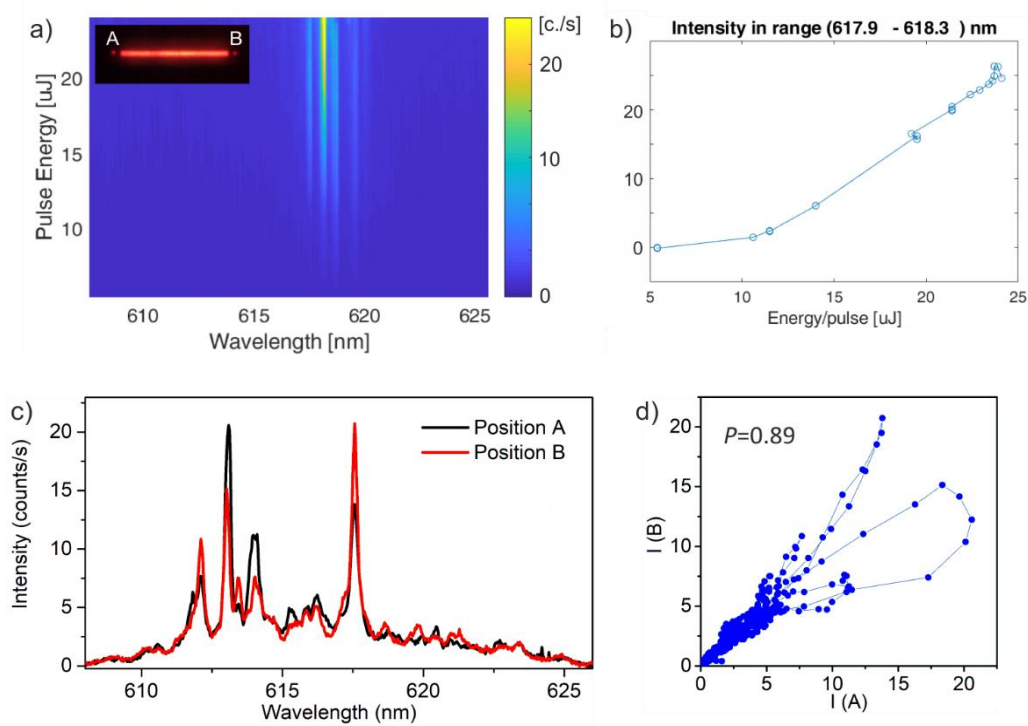

Figure S1. a) Spectra acquired for a single RL at one scattering centre position (A in the inset) as a function of the pumping-laser energy per pulse. b) Intensity integrated over the spectral range corresponding to the most intense RL peak, reported as a function of the energy per pulse delivered. c) Single RL spectra collected at the two scattering nodes A (black line) and B (red line), respectively, of a device with the same design of the one in a). d) Parametric plot (connected blue dots) of the intensity of the spectra in c) exhibits a linear correlation $P$=0.89.

[1]Instituto de Ciencia de Materiales de Madrid (ICMM), Consejo Superior de Investigaciones Científicas (CSIC), Calle Sor Juana Inés de la Cruz, 3, 28049 Madrid, Spain. [2]ETSI de Telecomunicación, Universidad Rey Juan Carlos, Calle Tulipán, 28933 Madrid, Spain [†]Current address: Dep. Química Física, Universidad Complutense de Madrid, Avenida Complutense, 28040 Madrid, Spain. *e-mail: n.caselli@csic.es, [#]e-mail: antonio.consoli@urjc.es [§]email: c.lopez@csic.es

Then, by acquiring spectra as a function of the exciting laser energy, the RL threshold can be estimated. In Fig.S1a) the spectra of a single RL as a function of the pumping energy are reported. Fig. 1b) exhibits a clear threshold behavior (around 10 µJ/pulse along the stripe) and, notably, the persistence of peaks occurring at the same wavelength across the entire energy range. In all the experiments described in the paper the energy of the pumping laser was set to a reference value (22 µJ/pulse for each stripe) in order to assure that the excitation was settled above threshold. We also proved that the delivered energy was low enough in order not to burn or deteriorate the active material. Each spectrum was acquired over 50 consecutive laser pulses (10 ns of duration and 10 Hz repetition rate) to average the intensity fluctuations of the pump laser. Therefore, every sharp spectral peak occurring at random frequencies identifies a RL mode. The spectral random distribution of the peaks assure that they do not belong to family of longitudinal Fabry-Perot modes.

In fact, the investigated devices are too long (L=2 mm, resulting in a free spectral range $\Delta\lambda = \frac{\lambda^2}{2nL} = 0.07$ nm) to sustain longitudinal modes with a spectral separation larger than the available resolution [1].

The RL emission emerging from the two opposite scattering centers (A and B in Fig. S1) of a given device showed and a strong correlation, as observed in similar devices [2]. This correlation is analogous to laser operating in a resonant Fabry-Perot cavity, with the difference that in a RL the mirrors are made of disordered materials, hence they introduce an arbitrary phase and amplitude modulation for each interacting light wave depending on the wavelength. In both spectra of Fig.S1c) the predominant peaks occurred at the same wavelength, indicating the presence of the same amplified modes. The small dissimilarities can be ascribed to the different behavior of the two output couplers, which scattered light in the normal direction to the sample plane, thus allowing the emission to be detected by our setup. The parametric plot reporting the intensity of one spectrum as a function of the intensity of the second one for each wavelength is shown in Fig.S1d). It highlights the strong correlation between the two spectra, being the Pearson linear correlation (P) equal to 0.89. We performed similar tasks on spectra emerging from the opposite scattering centers in many different RL realizations and we did find a similar high correlation in every tested case. Therefore, in order to characterize the emission of a given RL, in the manuscript we reported only the spectrum collected at a given output coupler.

We investigated the stability of the modes supported in a single RL at fixed pumping energy and at a given detection position, by acquiring many successive spectra as a function of time. In Figure S2a) are reported 30 successive spectra corresponding to an overall measure of 2.5 min involving 1500 pumping laser pulses. The map exhibits small intensity fluctuations. In order to quantify the stability of the random modes we evaluated the probability density map of the sum of the parametric plot emerging from of all the possible couples of spectra, reported in Fig. S2d). We obtained an average Pearson correlation of $<P>=0.98$ with a standard deviation of $\sigma_{<P>}=0.01$. This proves that the high correlation was maintained throughout the entire acquisition. In order to check the limit of the stability, we kept measuring up to 500 spectra.

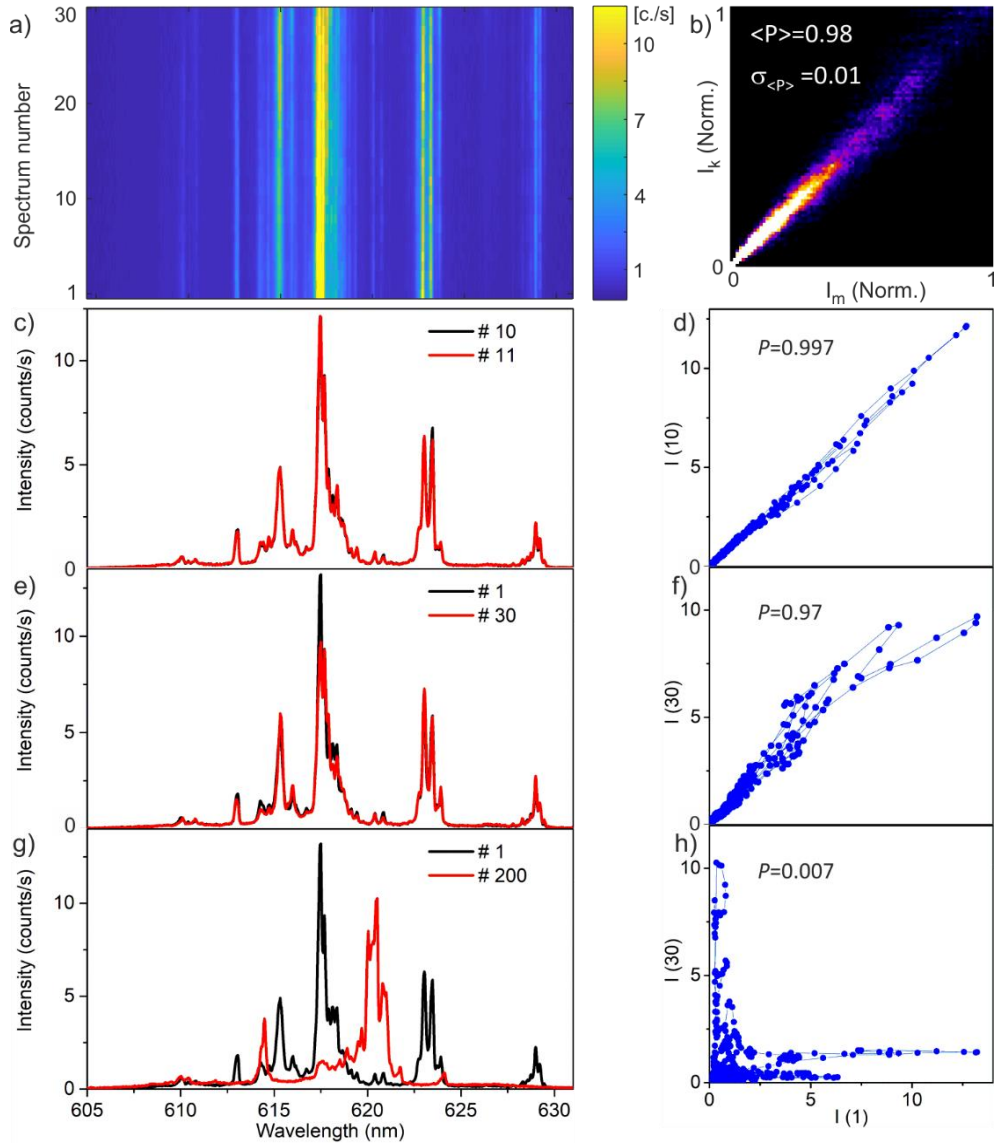

Figure S2. Stability analysis of a single RL emission. a) Successive spectra acquired at a given position as a function of time. Up to 30 successive spectra are reported, corresponding to an overall measure of 2.5 min and 1500 pumping laser pulses. b) Probability density map of the sum of the parametric plots evaluated between all the possible couple of different spectra reported in a). The indexes k,m run from 1 to 30 with k>m. The mean Pearson correlation between all the possible couple of spectra is $P$=(0.98±0.01). c)-d) Consecutive spectra (#10 vs #11) and the resulting parametric plot, respectively. e)-f) Spectra #1 vs #30 and the corresponding parametric plot, respectively. g)-h) Spectra #1 vs #200 (separated by $10^4$ shots) and the corresponding parametric plot, respectively.

Figure S2c)-h) shows the comparison between spectra acquired at different time delay. Two successive spectra, Fig. S2c)-d), give rise to a parametric plot with a narrow line distribution over the diagonal, indicating a strong correlation ($P$=0.997). The comparison between the first and the last spectrum of Fig. S2a), reported in Fig. S2e)-f), also results in a high correlation ($P$=0.97) due to minimal intensity fluctuations. By increasing the delay time, spectra began to lose correlation with the first one. A vivid example emerges from the comparison with the spectrum #200, Fig. S2g)-h), in which after about 17 min and $10^4$ shots the response presents negligible correlation with the initial one ($P$=0.007). Notably, in the spectrum #200 many RL modes were still amplified but with uncorrelated spectral position and different separation. The parametric plot is distributed out of the diagonal, mainly along the vertical and horizontal axes, indicating that the high intensity peaks of one spectrum correspond to spectral regions where the other spectrum presents low intensity.

## 2. Parametric plot analysis

For every couple of spectra, we constructed the so called "parametric plot" by reporting the intensity of the first spectrum as a function of the intensity of the second one, for each acquired wavelength. The line that connects the scatter points are drawn following the wavelength order. It visually highlights the correlation between the two spectra considered. In order to estimate the degree of similarity between spectra we evaluated the linear coefficient of correlation (Pearson coefficient, $P$, that is proportional to the variance of the angular distribution of the parametric plot data), and, in addition, the area ($S$) that lies in between the parametric plot lines. To calculate this parameter, each parametric plot was converted in a figure and the percentage of internal area was then counted by image processing. The conjunction of the two parameters $P$ and $S$ allowed us to estimate a degree of similarity exploited in finding RLs interaction in a network. In the case of almost identical spectra, the parametric plot presents a distribution of points squeezed to the diagonal, resulting in $P{\sim}1$ and $S{\sim}0$, see Fig. S2d). The case of two independent spectra is represented by peaks with small or negligible overlap, that generate a parametric plot distributed along the axes and with both $P{\sim}0$ and $S{\sim}0$, see Fig. S2h). In these two cases the linear correlation alone represents a good parameter to distinguish the emission behaviour, but what happen to intermediate values of P? What are the clues of an effective interaction between RLs?

To answer these questions, we calculated $P$ and $S$ for spectra consisting of a given number of Lorentzian peaks with the addition of a certain level of noise. The analysis of the corresponding parametric plots allowed us to benchmark the proposed correlation parameters on test-bed cases. For instance, in Fig. S3a)-f) are reported parametric plots corresponding to spectra exhibiting a single Lorentzian peak (with same amplitude and width) as a function of detuning, i.e. for increasing spectral overlap. The Pearson coefficient increases monotonically as the spectral overlap increases. However, the area $S$ reaches a maximum value for an intermediate overlap (detuning~width) and approaches zero in both cases of negligible and maximum superposition. Therefore, $S$ highlights the presence of a partial overlap in which, significantly, small deviations need to occur. Since in considering two RLs that share a common scattering center we are interested in evaluating an overlap that possesses a gentle kind of spectral and amplitude deviations, therefore the area $S$ can highlight an effective interaction between RLs.

Further notable examples of parametric plot calculations for spectra involving a larger number of Lorentzian peaks are reported in Fig. S3g)-l). For instance, in Fig. S3g) a spectrum with two separated peaks is compared to a spectrum with only one slightly shifted peak. In this case the overlapping (but not identical) peaks give rise to the larger contribution to $S$ and the isolated peak results in a vertical line with negligible area contribution. Notably, the case discussed in Fig. S3h) corresponds to the Rabi splitting that occurs for two resonators in the strong coupling regime. The single peak (blue line) represents the spectra of identical uncoupled resonators and the double peak (red line) the spectrum of the strong-coupled case. The single peak has a minimum overlap with the double peak spectrum, being exactly in the middle. The parametric plot displays a closed curve with some extent of correlation that do not possess a linear contribution ($P$=0.12), neither an area significantly larger than zero ($S$=0.01). However, since the strong coupling regime is not expected in RLs, the correlation underneath this case does not represent the case we aim to investigate. Therefore, the conjunction of $P$ and $S$ allow us to identify only weak coupling in RL networks. It is worth stressing the fact that parametric plots with a similar area can show a different $P$, as occurs for the cases of Fig. S3i)-l). This finding reinforces the reliability of the area parameter for proving the mutual influence of RL in cases of intermediate $P$.

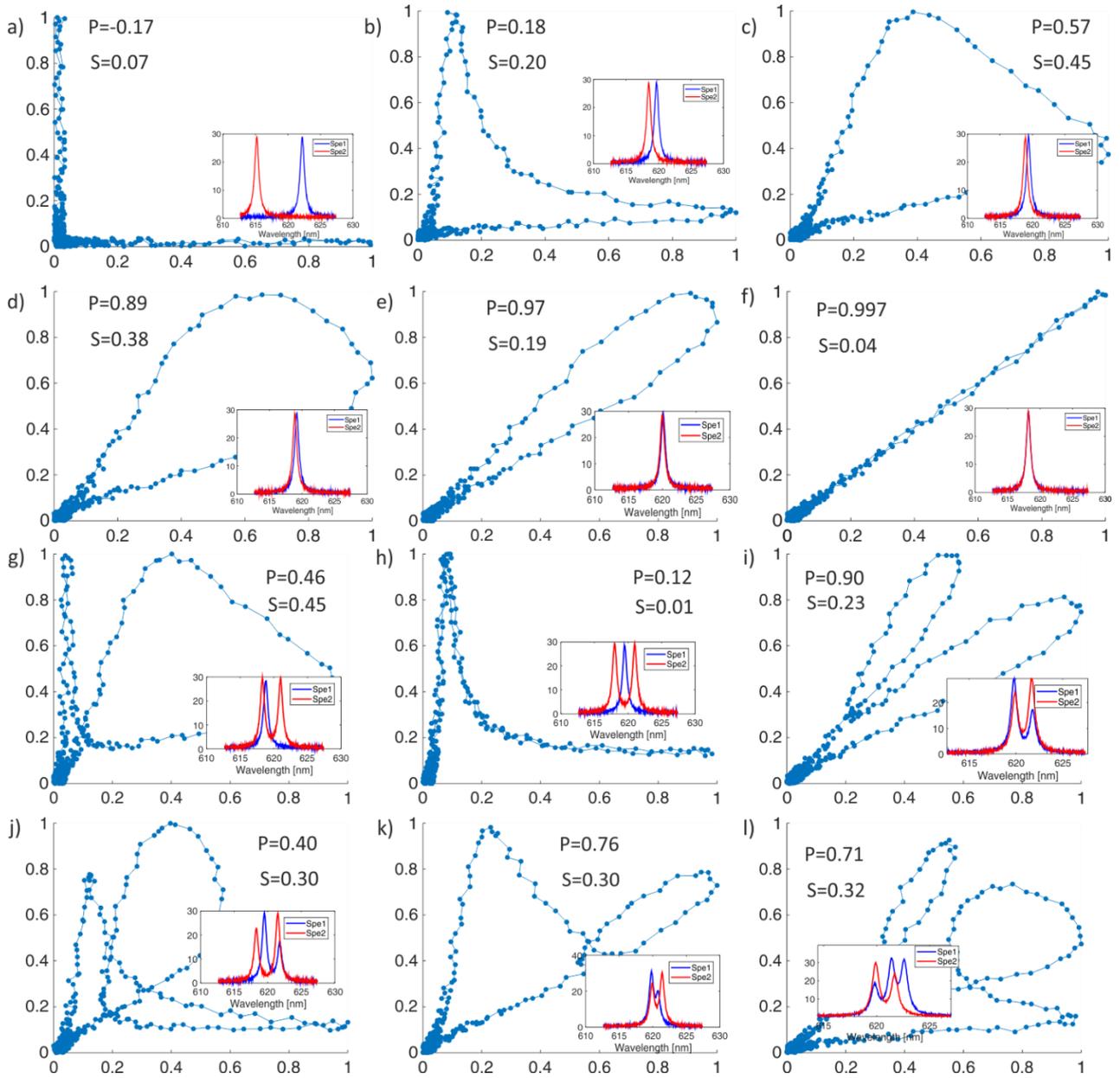

Figure S3. Parametric plots generated by two spectra (see insets) each composed of one Lorentzian peak (with same amplitude and width) reported for decreasing spectral separation (detuning). From a) to f) the detuning grows from values larger than the width (a) to almost zero (f), thus increasing the spectral overlap of the peaks. g)-h) Parametric plots generated by three total Lorentzian peaks. i)-k) Four total Lorentzian peaks. l) Five total Lorentzian peaks. The values of Pearson coefficient (*P*) and the area parameter (*S*) are reported for each case.

In order to summarize the results obtained by the calculations reported in Fig. S3 (and even more calculated spectra) the values of *S* and *P* are reported in Fig. S4a) as a function of the corresponding detuning/width for the single Lorentzian cases. While in Fig. S4b) *S* is reported as a function of Pearson coefficient. This proves that for intermediate values of *P* can be found cases with very different area parameter (ranging from 0.23 to 0.46). Therefore, the evaluation of *S* helped us to identify the cases in which spectral overlap is the fingerprint of an effective interaction between RLs.

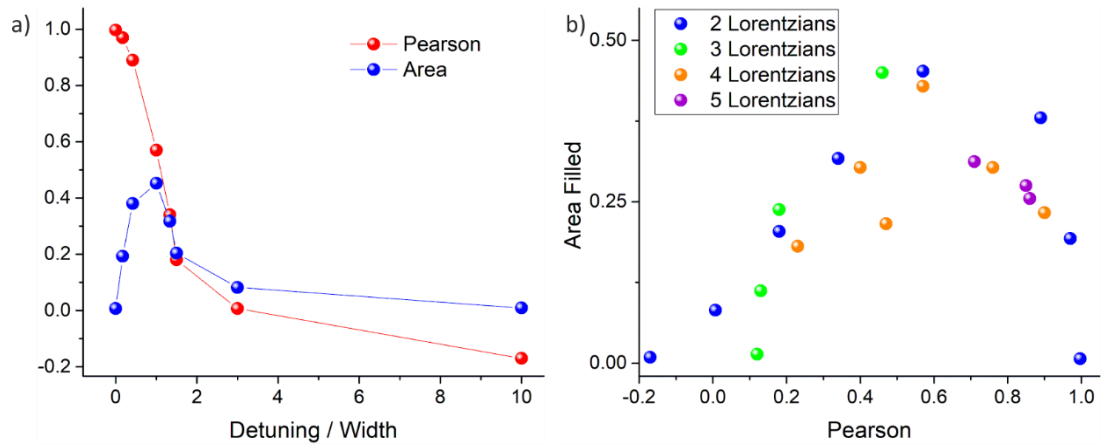

Figure S4. a) Pearson coefficient (red) and area parameter (blue) evaluated for spectra made of a single Lorentzian peak (Fig. S3) as a function of the spectral separation (reported as detuning over width of the Lorentzian peak). b) Area parameter as a function of Pearson coefficient for the spectra reported in Fig. S3 and similar ones. Different colours refer to the total number of Lorentzian peaks considered.

# 3. Random laser networks

In this section we discuss further data concerning RL networks presented in the main manuscript, but also samples with different scattering centers in order to corroborate our findings.

When investigating a network of two RLs, due to the output coupler difference, we chose a given scattering center position and then we acquired the spectra by pumping alternatively the first RL device, the second one and finally both of them at the same time, while keeping the detecting optical fiber in the same position.

In Fig. S5 are reported the Pearson coefficient and the area parameter ($S$) for 20 different network devices with the same design as the one reported in Fig. 2-3 of the main manuscript. This summary highlights the different correlations emerging when comparing two independent RLs (1vs2) and the sum of the two RLs with the compound (1+2 vs 1&2). By means of the correlation analysis we found fingerprint of coupling ($S >$ 0.15 and $P <$ 0.9) in 40% of the cases; while the interaction was poorly noticeable in the other cases (20% of cases with 0.10 < $S$ < 0.15; 40% with $S$ ≤ 0.10). Moreover, for the 20 sets of independent RLs we found a mean area parameter < $S$ >=0.05 with $\sigma_S$=0.03, and a maximum value $S_{Max} = 0.13$. Then, in order to define a proper threshold for interaction between RLs, $S_{Th} = 0.15$ should be effective in highlighting cases in which the modes of different RL are coupled.

By investigating devices with different geometries (size and distribution of the single holes), although they showed a different distribution in the number of modes, we found the overmentioned statistics to be independent on the scattering center design.

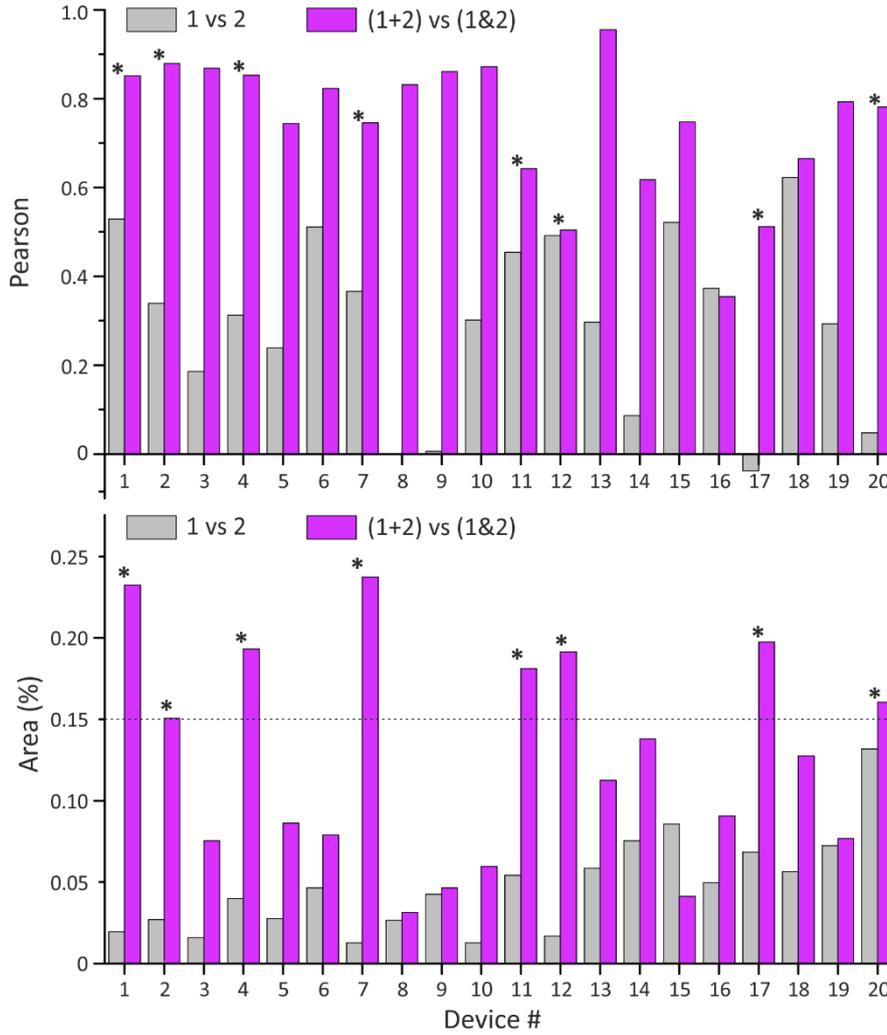

Figure S5. Correlation in 20 networks made of 2 RLs sharing a scattering centre. The Pearson coefficient (top) and the area parameter (bottom) are reported for the two independent RLs (1vs2, gray bars) and for the comparison between the sum of the two RLs with the compound (1+2 vs 1&2, purple bars). The (*) symbol highlights the devices where an effective coupling was demonstrated by correlation analysis.

In Fig. S6 we report a photonic network with three nodes positioned at the vertexes of an equilateral triangle with (side length of 2.5 mm), each scattering center was made of a single hole with diameter size around 150 μm and rough internal surface, see the inset of Figure S6 a). These scatterers were larger than the ones presented in the manuscript, however they provided a similar feedback to trigger RL action (even if the collected intensity was significantly larger due to the increased out of plane coupling efficiency of the bigger scatterers). Two independent RLs were induced by pumping lines connecting the vertexes A-B (line 1) and B-C (line 2), respectively, as shown in the insets of Figure S6 a). The low correlation is highlighted by the parametric plot, Figure S6 b), which exhibits $P$=0.08 and $S$=0.13, hence proving the independence of the two single RLs. The area value was close to the threshold $S_{th}$=0.15 previously defined for independent spectra. The non-zero $S$ value was possibly given by the contribution of only two single random modes that partially overlaps. This accidental overlap is not surprising, since, in the spectral range in which amplified spontaneous emission of the dye occurs, random modes can emerge stochastically at any wavelength. However, being $S<S_{th}$, this finding corroborated our thesis that, for intermediate $P$ values, an effective RLs interaction is evidenced by an area parameter larger than $S_{th}$.

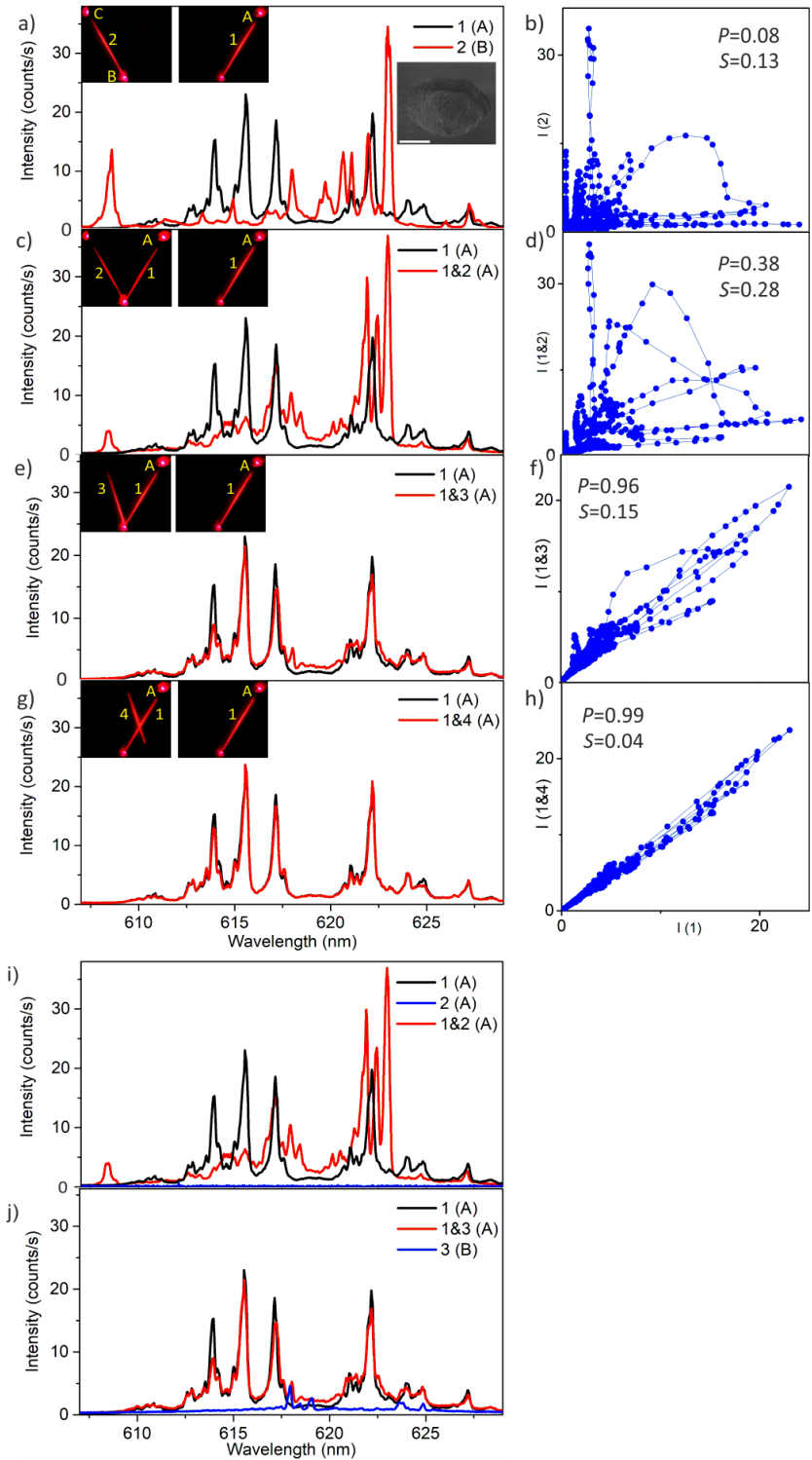

Figure S6. RL network for scattering centres made of single holes. a) Spectrum acquired in position A by pumping line 1 (black curve) and in position B by pumping line 2 (red curve). In the insets the out-of-plane emission along with the SEM image of a scattering centre (scale bar 50 μm). b) Parametric plot of the spectra reported in a). c) Spectra acquired in position A by pumping line 1 (black curve), line 1 and 2 simultaneously (1&2, red curve). d) Parametric plot of the spectra reported in c). e) Spectra obtained in A by pumping line 1 (black curve), line 1 and 3 simultaneously (1&3, red curve). f) Parametric plot of the spectra reported in e). g) Spectra obtained in A by pumping line 1 (black curve), line 1 and 4 simultaneously (1&4, red curve). h) Parametric plot of the intensity of the spectra reported in g). The values of Pearson correlation (*P*) and of the area of the parametric plot (*S*) are reported for each case. i) Spectra detected in position A by pumping RL 1 only (black curve), RL 2 only (blue line), RL 1 and 2 simultaneously (red curve). j) Spectra acquired in position B by pumping line 3 only (blue line); in position A by pumping RL 1 only (black curve); in position A by pumping RL 1 and line 3 simultaneously (red curve).

We evaluated the spectral modifications induced on the emission of RL 1 by the simultaneous presence of the RL 2, as reported in Fig. S6c)-d). The Pearson coefficient increases to $P=0.38$ and the area of the parametric plot to $S=0.28$. If we analyze in details the spectrum measured when both lines are pumped, a variety of mode behaviors can be found: *i)* the persistence of peaks owned by RL 1; *ii)* the introduction of new modes, which belong to the stand-alone RL 2, appearing in position A due to the enhancement along the stripe that connects B to A; *iii)* the modification RL 1 resonances, giving rise to peaks with slightly different wavelength and intensity, showing a partial overlap with the emission of the single devices. All these behaviors are well discriminated in the parametric plot, since diagonal lines with small area correspond to case *i)*, both vertical and horizontal lines with small area are ascribed to case *ii)*, while curves that induce larger areas emerge from case *iii)*. In summary, the linear correlation proves that two spectra are independent ($P<0.15$) or strongly correlated ($P>0.9$), while for intermediate $P$ values the evaluation of the area parameter is able to discern the modification on the emission if $S > S_{th}$, and, therefore, the presence of an effective RLs coupling.

The power of the method is strengthened when investigating cases in which the coupling is absent, for instance the pumping schemes reported in Figure S6 e)-h), since negligible modifications were induced on RL 1. In the first case, Fig. S6 e)-f), line 3 shared a scattering center with RL 1 but did not have another disordered mirror at the opposite end, so the stand-alone device #3 does not support strong RL resonances, as reported in Fig. S6 j). The modifications that line 3 induced on the emission of RL 1 were quite negligible and the output parameters ($P=0.96$, $S=0.15$) proved the absence of an effective interaction. In the second case, Fig. S6g)-h), line 4 shared no scattering center with RL 1 and induced even a lower modification to the spectrum observed in A if compared to the one generated by line 3 ($P=0.99$ and $S=0.04$).

Fig. S6 i) shows that when RL 1 is off, no emission from RL 2 is detected in position A (blue flat line). Therefore, the peaks introduced by pumping both lines are ascribed to the amplification that took place inside the line that connects A to B (peaks in the red spectrum that are not present in the black one) or to the coupling between the two RLs (peaks in the red spectrum exhibiting slightly changes in intensity and/or resonant wavelength with respect to the black one). Interestingly, the spectrum detected in B by pumping the sample by the stand-alone line 3 shows very low intensity peaks, as reported in Fig. S6 j). In fact, due to the absence of a scattering center at the end of line 3, a feedback element is missing, however the presence of inhomogeneity in the dye-polymer along the pumping stripe can act a scatterers and induce small intensity resonances. It is noteworthy to stress that the peaks of line 3 detected in B correspond to the ones observed in position A when both RL 1 and line 3 were simultaneously active, thus indicating the small interaction between the two devices.

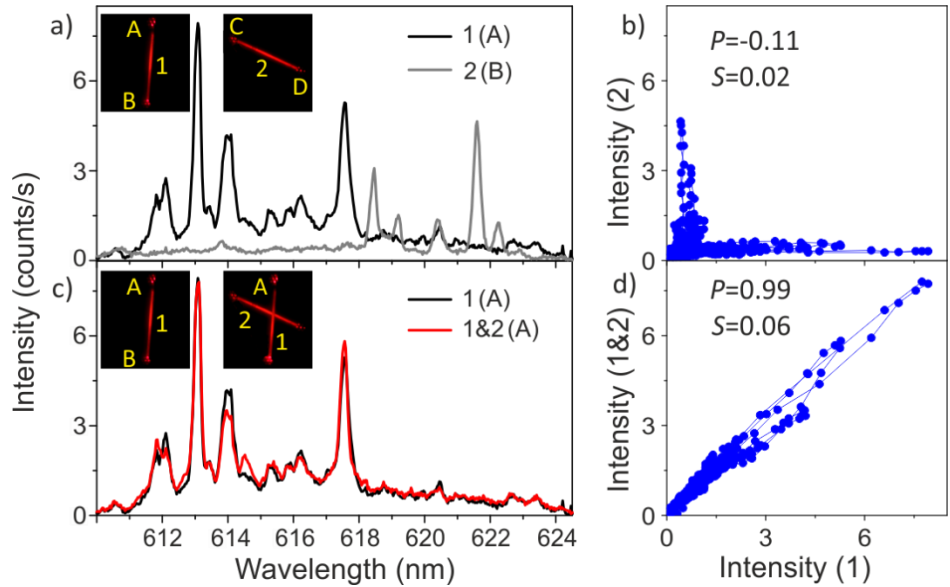

Figure S7. Two RLs that are not end-coupled. a)-b) Spectra and parametric plot, respectively, of the two independent RLs: the one connecting the scattering centres A-B (RL1, black line) and the one connecting the scatting centres C-D (RL2, grey line). In the inset the out-of-plane emission of the network is reported. c)-d) Spectra and parametric plot, respectively, of RL 1 (black line) and of the compound of both RLs pumped simultaneously (1&2, red line), acquired in position A. The values of Pearson correlation (*P*) and of the area of the parametric plot (*S*) are reported for each case and demonstrate the absence of coupling in the geometry in which the RLs are not end-coupled.

In Figure S8 we reported further details regarding the emission stability of the RL ring network presented in Figure 5 of the manuscript. By analyzing the spectral response at the three nodes (A,B,C in the inset) detected when all the three RLs were simultaneously active, we proved the high correlation among them. In fact, by comparing the spectra couple by couple, a high Pearson coefficient *P*>0.9 and small area parameter *S*<0.1 were found in all the cases. Although the spectra showed intensity variation, remarkably, the peak positions remained almost unchanged and most of the peaks displayed an intensity variation dependent on their spectral position with respect to the maximum dye emission. Therefore, the observed differences clarify that they were not independent resonators. These fluctuations could be due to unavoidable fabrication difference in the network nodes, which scatter the RL emission out of the sample plane in a different way, thus giving rise to intensity variation without losing correlation. Moreover, the absence of a spectral rearrangement or of a resonance spectral shift, marked a huge contrast with the coupled RLs cases presented in this work, in which the parametric plot showed a lower *P* and a higher *S*. Finally, this spectral analysis demonstrates that the compound RL modes are detected at the three nodes of the network, as it was a single large resonator.

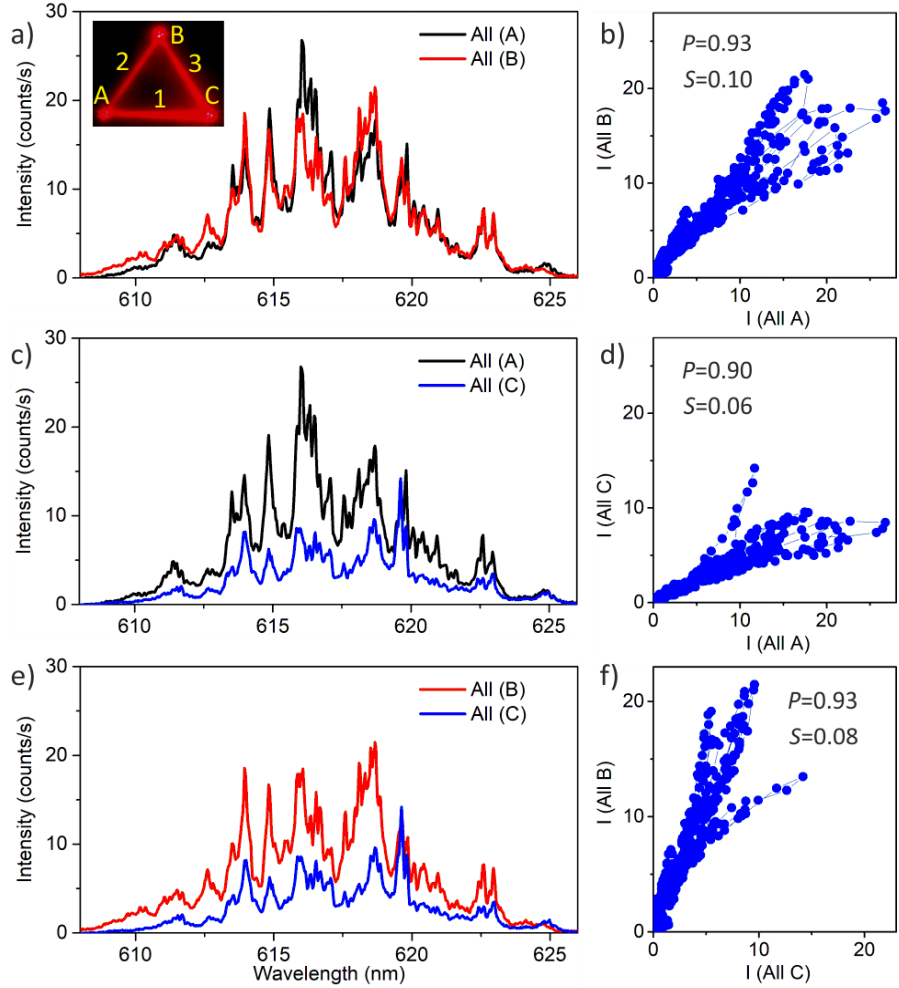

Figure S8. Ring network of three RLs (1,2,3) investigated at the three nodes (A,B,C) when all the 3 RLs are simultaneously pumped. a)-b) Spectra and parametric plot, respectively, of the network emission in position A (black line) and B (red line). In the inset the out-of-plane emission of the network is reported. c)-d) Spectra and parametric plot, respectively, of the network emission in position A (black line) and C (blue line). e)-f) Spectra and parametric plot, respectively, of the network emission in position B (red line) and C (blue line). The values of Pearson correlation ($P$) and of the area of the parametric plot ($S$) are reported for each case.

In conclusion, the interaction between two nominally identical RLs cannot be predicted a priori, as the random nature (concerning the spectral and spatial distribution) of the modes overlapping on the common disordered scattering center prevents any actual device to be modeled. However, in our study we highlighted the occurrence of RLs in which the interaction can be quantified by the variation of the $P$ and $S$ parameters with respect to the uncoupled case. In comparison to other methods that evaluate the similarity between different spectra, such as the autocorrelation function or the mutual information, our approach gives a remarkable outcome in highlighting spectral redistributions. In fact, the summary of the experimental cases analyzed in the paper and in the SI, shown in Fig. S9, highlights how the area parameter as a function of $P$ has a sharp maximum for intermediate value of $P$, while the mutual information (evaluated as in Ref. [3])has a monotonic behavior that is not of any help in finding coupling signatures. On the other hand, the dissimilarity criterion based on Ref. [4] gives a linear correspondence with respect to the Pearson coefficient.

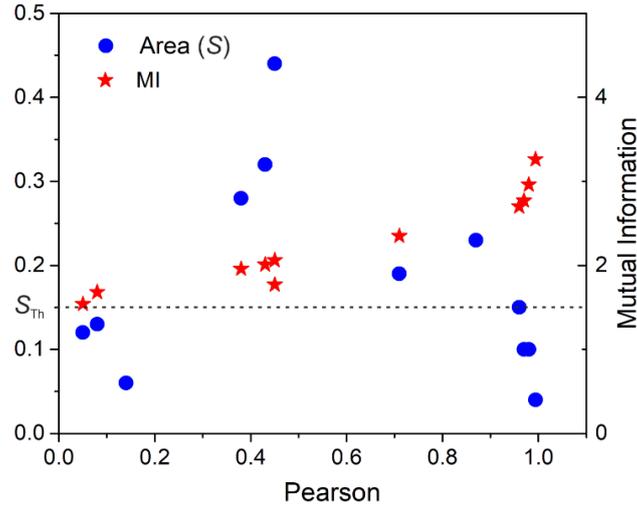

Figure S9. Values of the area parameter obtained for the experimental cases presented in this work (blue dots) and the corresponding value of the mutual information (MI) as a function of the Pearson coefficient ($P$). $S_{Th}$=0.15 corresponds to the threshold above which we consider that the spectral variation is due to an effective interaction between RLs in the network.

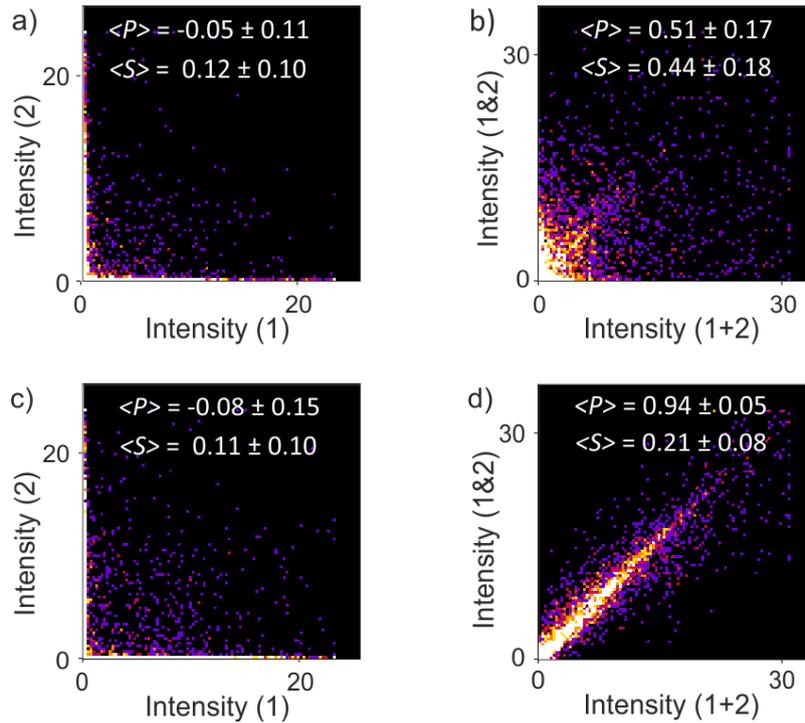

Figure S10. Numerical simulations employing coupled mode theory considering 20 different networks, each made of 2 RLs with 10 active modes. a) Probability density map of the sum of 20 parametric plots, each as the one reported in Figure 6b) of the main manuscript, evaluated for 20 simulations with independent sets of modes (RL 1 vs RL 2). b) Probability density map of the sum of 20 parametric plots in which the sum of the two spectra (1+2) analyzed in a) was compared to the compound (1&2) in the case of the same coupling strength set in the calculations of Fig.6 of the manuscript: $c_{max} = 1.5 \cdot 10^{-5}$ ($g_{max} - \alpha_{max}$), where $g_{max}$ and $\alpha_{max}$ correspond to the maximum gain and losses, respectively. c) Probability density map of the sum of 20 parametric plots evaluated for 20 simulations with independent sets of modes (RL 1 vs RL 2), as reported in a) but using a different set of random modes. d) Probability density map of the sum of 20 parametric plots in which the sum of the two spectra (1+2) analyzed in c) was compared to the compound (1&2) for reduced coupling strength in comparison to b): $c_{max} = 0.75 \cdot 10^{-5}$ ($g_{max} - \alpha_{max}$). The mean values of the Person correlation $<P>$ and of the area parameter $<S>$ are reported for each case along with the corresponding standard

deviations. In a) and c) the low values of $< P >$ and $< S >$ proved the absence of correlation between the independent set of spectra 1 vs 2. When comparing (1+2) to (1&2), the case of larger coupling, b), gave rise to a wider cloud distribution in the map with respect to d), where most of the data are distributed along the diagonal. In b) the intermediate value of $< P >= (0.51 \pm 0.17)$ and the high $< S >= (0.44 \pm 0.18)$ are the fingerprint of a strong interaction between the 2 RLs considered, while in d) the high linear correlation $< P >= (0.94 \pm 0.05)$ and the decreased area parameter $< S >= (0.21 \pm 0.08)$ with respect to b), demonstrated the strong similarity between the sum and compound spectra, fingerprint of a negligible RL interaction.